\documentclass[preprint,journal]{vgtc}  

\ifpdf
  \pdfoutput=1\relax                   
  \usepackage{graphicx}                
  \DeclareGraphicsExtensions{.pdf,.png,.jpg,.jpeg} 
\else
  \usepackage{graphicx}                
  \DeclareGraphicsExtensions{.eps}     
\fi%

\graphicspath{{figures/}{pictures/}{images/}{./}} 

\usepackage{microtype}                 
\PassOptionsToPackage{warn}{textcomp}  
\usepackage{textcomp}                  
\usepackage{mathptmx}                  
\usepackage{times}                     
\usepackage{cite}                      
\usepackage{tabu}                      
\usepackage{booktabs}                  

\vgtccategory{Research}
\vgtcpapertype{Application/Design Study}
\ieeedoi{10.1109/TVCG.2019.2934614}
\title{Interactive Learning for Identifying Relevant Tweets to Support Real-time Situational Awareness}

\author{Luke~S.~Snyder, Yi-Shan~Lin, Morteza~Karimzadeh, Dan~Goldwasser, and David~S.~Ebert,~\textit{Fellow,~IEEE}}
\authorfooter{
\item
 Luke S. Snyder, Yi-Shan Lin, Dan Goldwasser, and David S. Ebert are with Purdue University.
 \\Email: \{snyde238, lin670, dgoldwas, ebertd\}@purdue.edu
 \item Morteza Karimzadeh is with the University of Colorado Boulder (formerly at Purdue University). Email: karimzadeh@colorado.edu
}

\shortauthortitle{Biv \MakeLowercase{\textit{et al.}}: Global Illumination for Fun and Profit}

\usepackage[dvipsnames]{xcolor}
\usepackage{soul}
\usepackage{url}
\usepackage{floatrow}
\colorlet{red}{black}

\abstract{
Various domain users are increasingly leveraging real-time social media data to gain rapid situational awareness.
However, due to the high noise in the deluge of data, effectively determining semantically relevant information can be difficult, further complicated by the changing definition of relevancy by each end user for different events.
The majority of existing methods for short text relevance classification fail to incorporate users' knowledge into the classification process. Existing methods that incorporate interactive user feedback focus on historical datasets.
Therefore, classifiers cannot be interactively retrained for specific events or user-dependent needs in real-time.
This limits real-time situational awareness, as streaming data that is incorrectly classified cannot be corrected immediately, permitting the possibility for important incoming data to be incorrectly classified as well.
We present a novel interactive learning framework to improve the classification process in which the user iteratively corrects the relevancy of tweets in real-time to train the classification model on-the-fly for immediate predictive improvements.
We computationally evaluate our classification model adapted to learn at interactive rates. Our results show that our approach outperforms state-of-the-art machine learning models. 
In addition, we integrate our framework with the extended Social Media Analytics and Reporting Toolkit (SMART) 2.0 system, allowing the use of our interactive learning framework within a visual analytics system tailored for real-time situational awareness.
To demonstrate our framework's effectiveness, we provide domain expert feedback from first responders who used the extended SMART 2.0 system.
} 

\keywords{Interactive machine learning, human-computer interaction, social media analytics, emergency/disaster management, situational awareness}

\teaser{
  \centering
  \includegraphics[width=\linewidth]{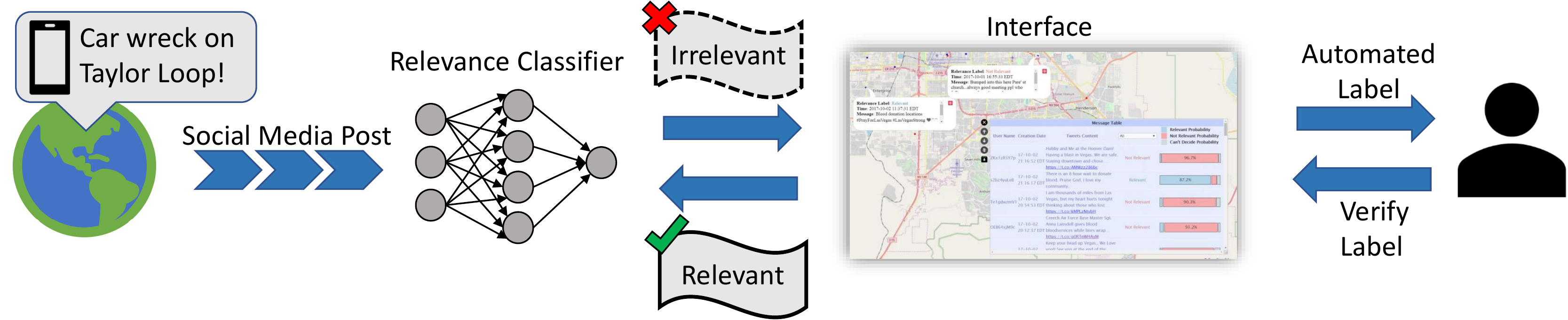}
  \caption{Our interactive learning framework allows users to train text relevance classifiers in real-time to improve situational awareness. In this example, a real-time tweet regarding a car accident is incorrectly classified as ``Irrelevant''. Through the SMART 2.0 interface, the user can view its label and correct it to ``Relevant'', thereby retraining and improving the classifier for incoming streaming data.}
  \label{fig:teaser}
}

\vgtcinsertpkg

\begin{document}



\maketitle

\section{Introduction}
Social media data has been used extensively in a variety of applications and research endeavors due to its ability to provide useful information on the public's opinions and behavior.
Analysts in various domains are increasingly using social media to gain rapid situational awareness. For instance, first responders are leveraging Twitter data to obtain actionable information for crisis response and prevention (see~\cite{MartíNez-RojasMaría2018Taat} for an extensive list of literature on this subject). 
However, the vast amounts of  unstructured text make the identification of relevant information nontrivial, limiting situational awareness.
This issue is further compounded by changes in topics of interest (to end users) over time, since the computational models built to determine relevant information for one event or one user group may not apply to other events or other user groups due to variations in diction, word structure, or user expectations. 

Several classification approaches have been developed to identify relevant and irrelevant social media information, such as clustering~\cite{BeckerHila2011BTTR, AshktorabZ.2014TMtt}, keyword matching~\cite{ToHien2017OIDT}, and term-vector similarity~\cite{DiakopoulosN2010Ditr}.
However, to the best of our knowledge, no existing work in this area includes interactive learning with real-time data, focusing instead on improving the machine learning algorithms themselves~\cite{AshktorabZ.2014TMtt, ghosh2018class, KarimiSarvnaz2013Cmfd, NazerTahoraH2016Fris, NguyenDatTien2016AoOD, NguyenDatTien2016RCoC, RudraKoustav2015ESIf, ToHien2017OIDT, toriumi2016real, ZoppiTommaso2018Lret} or interactively training on archived datasets~\cite{BoschHarald2013SRMo, HeimerlF2012VCTf}.
Continuing on our example of first responders, a pre-trained classifier may not fulfill first responders' varying needs, since one first responder may be interested in monitoring road closures, and another one might be interested in identifying disinformation and misinformation on social media in order to take counter-action. 
Ultimately, first responders' definition of relevancy will depend on the situation at hand, which can vary over time.
Interactively training classifiers through iterative user labeling can alleviate this problem.

In this paper, we present a novel interactive framework in which the user iteratively (re)labels the relevancy of streaming social media data to adaptively train the underlying model to match their needs for improved situational awareness. We compare three different types of neural networks in terms of classification performance and computational efficiency for real-time learning. 
Furthermore, we optimize and computationally evaluate the selected models by simulating the real-time user feedback on several crisis-related datasets. Our results show that our interactive model outperforms state-of-the-art machine learning-based classification models.

To incorporate our evaluated models into a working application, we extend an existing visual analytics system tailored for situational awareness called the Social Media Analytics and Reporting Toolkit (SMART)~\cite{zhang2017smart, ZhangJiawei2014Riam}, which has been successfully used by many first responder groups in the United States. SMART allows users to interactively explore trending topics on social media through integrated topic modeling and spatial, temporal, and textual visualizations. We call the newly extended system SMART 2.0, which incorporates our interactive learning framework to address the needs raised by the aforementioned first responder users and reduce noise in the incoming stream of data.

Finally, we present domain-expert feedback on the usefulness of our approach as experienced by multiple first responders who used SMART 2.0 for crisis-related use cases.
In addition, we include two usage scenarios of the system to illustrate its application to real-life situations.
Overall, the major contributions of this paper are as follows:
\begin{enumerate}
    \item We present a novel interactive learning framework for  classification of streaming text data.
    \item We compare three different types of neural networks in terms of performance and computational efficiency, and tune the models for learning at interactive rates. We further computationally evaluate the selected model on several disaster-related datasets.
    \item We integrate our models in SMART 2.0, a visual analytics application for situational awareness, and present user feedback obtained from domain experts using the system for crisis events.
\end{enumerate}

In the remainder of the paper, we discuss related work in section 2, the design of the framework and model in section 3, SMART 2.0 in section 4, evaluation of our framework in section 5, discussion and future work in section 6, and concluding remarks in section 7.

\section{Related Work}
\subsection{Short Text Classification}

Researchers have presented many techniques to classify text documents into categories such as sentiment or topics~\cite{WilsonTheresa2009RCPA, go2009twitter, pennacchiotti2011machine, PhanX.-H.2008Ltcs}. 
However, classifying short text, e.g. social media posts, is more challenging due to the lack of contextual information and loose adherence to standard grammar. 
To tackle the brevity of short text, auxiliary resources such as external corpora~\cite{chen2011short} or knowledge bases~\cite{HuXia2009Eiae}, or methods such as term frequency-inverse document frequency (TF-IDF)~\cite{ghosh2018class}, have been proposed for improving classification.

Representing words as $n$-dimensional vectors (i.e. word embedding) has become increasingly prevalent, since vectors can be used as inputs to machine learning models for finding semantic similarities~\cite{WongPakC.2004II2C, TrieuLap2017NCfS}.
In particular, Google's \texttt{Word2Vec}~\cite{MikolovTomas2013DRoW} has been employed extensively in classification tasks~\cite{LingW.2015Tsao, LillebergJoseph2015Svma, abdelwahab2016uofl, NguyenDatTien2016RCoC, NguyenDatTien2016AoOD} due to its impressive ability in capturing linguistic regularities and semantics.
For instance, words frequently used together are likely to be closer in the \texttt{Word2vec} vector space than words that are not, and vector operations reveal meaningful semantics (e.g., the vector ``King'' $-$ ``Man'' $+$ ``Woman'' is close to the vector ``Queen''~\cite{MikolovTomas2013DRoW}).
Since pre-trained \texttt{Word2vec} models encode embeddings learned from larger web corpora, they have been increasingly used in short text classification tasks~\cite{abdelwahab2016uofl, NguyenDatTien2016AoOD, NguyenDatTien2016RCoC, TangD.2014Lswe}.

Neural networks have generated state-of-the-art results in recent years for text classification problems~\cite{NguyenDatTien2016AoOD, NguyenDatTien2016RCoC, TangD.2015Dmwg, MikolovTomas2013DRoW} and have also been used with \texttt{Word2Vec}~\cite{NguyenDatTien2016RCoC, NguyenDatTien2016AoOD, TangD.2014Lswe}. 
Neural networks are well-suited for online learning processes in which training data is supplied iteratively since they can learn adaptively from new data~\cite{NguyenDatTien2016AoOD, NguyenDatTien2016RCoC}.
Nguyen et al.~\cite{NguyenDatTien2016RCoC} presented a convolutional neural network with \texttt{Word2Vec} that outperformed non-neural classifiers, and Nguyen et al.~\cite{NguyenDatTien2016AoOD} proposed a new online learning classification algorithm for deep neural networks utilizing the log-loss and gradient of sequential training batches.
Their methods were evaluated with disaster-related datasets.
However, these methods were not adapted to user-guided learning in which time constraints are essential and the provided batches may be small. 
In particular, the online learning method designed by Nguyen et al.~\cite{NguyenDatTien2016AoOD} was evaluated with batch sizes of 200. 
In our work, we assume the user needs to train with flexibly interactive amounts of data (10-20 samples) to view immediate predictive improvements for situational awareness. 

\textbf{Classification for situational awareness.} Utilizing real-time social media data for situational awareness (and crisis prevention in particular) is a heavily researched topic~\cite{KarimiSarvnaz2013Cmfd, ghosh2018class, RudraKoustav2015ESIf, NazerTahoraH2016Fris, ZoppiTommaso2018Lret, ToHien2017OIDT, NguyenDatTien2016AoOD, NguyenDatTien2016RCoC}.
However, identifying situationally-relevant information is nontrivial due to the high noise-to-signal ratio.
Karimi et al.~\cite{KarimiSarvnaz2013Cmfd} found that classification methods, such as Support Vector Machine and multinomial Na\"ive Bayes, can identify disaster-related tweets, although generic features such as hashtag count and tweet length are preferable so that the model does not learn relevancy only for a specific disaster. 
Researchers have used clustering~\cite{BeckerHila2011BTTR, AshktorabZ.2014TMtt, toriumi2016real} or enhanced keyword matching~\cite{ToHien2017OIDT} to detect relevant crisis and event information, and provided human-annotated Twitter corpora that can be used to train word embedding models~\cite{ImranMuhammad2016TaaL}.

Nazer et al.~\cite{NazerTahoraH2016Fris} developed a system to detect requests for help by utilizing both tweet context (e.g., geotag) and content (e.g., URLs). Rudra et al.~\cite{RudraKoustav2015ESIf} designed a novel classification-summarization framework to classify disaster-related tweets, and then summarize the tweets by exploiting linguistic properties typical of disaster tweets (e.g., combinations of situational and non-situational information). Zoppi et al.~\cite{ZoppiTommaso2018Lret} provided a relevance labeling strategy for crisis management that computed data relevance as a function of the data's integrity (e.g., are the geo-coordinates incorrect?), statistical properties (e.g., can we select a subset of the data that are geographically close?), and clustering (e.g., what groups are present in the data?).
Toriumi et al.~\cite{toriumi2016real} clustered tweets based on their retweet count in real-time to extract important topics and classify tweets accordingly.

The methods discussed so far, however, lack user interactivity. In particular, these classification methods are inflexible to user-dependent needs that change over time as new situations and events occur. As such, their practical use for real-time situational awareness is limited.
\subsection{Visual Analytics and Interactive Learning for \\ Situational Awareness}
Researchers have presented a number of visual analytics (VA) solutions for situational awareness.
Diakopoulos et al.~\cite{DiakopoulosN2010Ditr} developed Vox Civitas, a VA application  for journalistic analysis and user-guided filtering using social media content.
Vox Civitas filters out unrelated data by automatically computing time-dependent term-vector similarities. 
TwitInfo~\cite{MarcusAdam2011Taav} aggregates streamed Twitter data and automatically discovers events from activity peaks in real-time. 
The authors assign relevance to a tweet by counting its number of event-related keywords.
Pezanowski et al.~\cite{PezanowskiScott2018Sagf} designed the geovisual analytics system SensePlace3 to provide situational awareness by leveraging geographical information and place-time-theme indexing with string-based queries for exploring datasets.
SensePlace3 primarily relies on TF-IDF for tweet retrieval in response to user queries. 
However, these tools do not employ machine learning for relevance classification and do not integrate user feedback to improve their underlying models or algorithms. 

Visual analytics has also been increasingly used to improve various machine learning processes, such as feature selection~\cite{dy2000visualization}, attribute weighting~\cite{WallEmily2018PRDU}, and labeling~\cite{BoschHarald2013SRMo, HeimerlF2012VCTf, BernardJ.2018CVLw}, and even understanding the models themselves~\cite{TalbotJustin2009Eivt, KahngM.2018AVEo, SachaDominik2019VAOf}. 
Sacha et al.~\cite{SachaD.2016Hmlt} proposed a framework to discuss the various forms of human interaction with machine learning models in visual analytics systems and theorized that VA tools could increase knowledge and usability of machine learning components. 
Endert et al.~\cite{EndertAlex2012Sifv} designed a system that classifies archived documents through user-guided semantic interactions (e.g., moving a document to another group) that improve the underlying model.
Our work is based on the same idea in that we intend to improve model performance through user feedback, but with real-time social media data. 

Heimerl et al.~\cite{HeimerlF2012VCTf} analyzed three separate methods for user-guided classification of a set of archived text documents: the \textit{basic} method, which does not employ sophisticated visuals; the \textit{visual} method, which visually represents the labeled and unlabeled documents for user exploration; and the \textit{user-driven} method, which provides the user with full control over the labeling process. The first two methods employ active learning, in which the model selects a data sample to be labeled by the user that most effectively helps it distinguish relevant from irrelevant data. This contrasts with the user deciding which instances they wish to label.
The authors did not find any statistically significant differences in terms of $F_1$ score between the methods in their user study. 
Bosch et al.~\cite{BoschHarald2013SRMo} developed ScatterBlogs2, a VA application that provides user-guided learning of filter classifiers on historical social media messages to support situational awareness. 
These two works are perhaps the most similar to ours, yet differ in two fundamental ways.
First, they do not provide interactive learning in real-time, which strains the user, as they are required to visit historical data for additional training. 
Second, they do not employ neural networks, which are better suited for online learning environments, such as social media streaming, in which training data is supplied sequentially over time~\cite{NguyenDatTien2016AoOD, NguyenDatTien2016RCoC}.
It is important to note that Bosch et al.~\cite{BoschHarald2013SRMo} allow the user to adjust a filter's focus (i.e., how precise the classification is) in real-time if it misses relevant data or does not sufficiently filter out irrelevant data.
However, this could indicate that the model has not properly learned the distinction between relevant and irrelevant data. Since training can only be completed with historical posts, the user is unable to update the model immediately with the streamed data, limiting situational awareness.
Our approach not only solves this issue by allowing the user to immediately train the model for improvement, but also provides the user with the ability to create classifiers on-the-fly to accommodate their real-time needs.

\section{Interactive Learning Framework}
Our framework for interactively learning relevant social media posts in real-time consists of two primary components. The first is a formalized set of design goals necessary to effectively facilitate situational awareness in real-time 
through user interactivity. The second is a detailed underlying model that is adapted to user-guided training with real-time streaming data. In section 4, we discuss our implementation of the framework that realizes the design goals.
\subsection{Design Goals}
The framework's design goals were iteratively defined through discussions with domain experts such as first responders who frequently use visual analytic social media applications for real-time situational awareness. In general, these experts found it necessary for the interactive framework to incorporate user feedback in a timely manner, as well as account for time and situation-dependent user needs. With their feedback, the following specific design goals were established:
\begin{enumerate}
    \item[\textbf{DG1}] \textbf{Filter and view relevant data:}
    Filtering data by relevancy removes noisy data, allowing the user to more quickly find data that may require immediate attention or contain important information. 
    The ability to view the relevant data itself is equally important for determining the urgency and content of relevant data. 
    
    \item[\textbf{DG2}] \textbf{Correct incorrect classifications:}
    Since classifiers may provide incorrect results, especially during the early stages of training, it is necessary for the user to be able to correct the label in real-time. This both improves the model's performance and lowers the likelihood that incoming streamed data will be incorrectly classified and missed.
    
    \item[\textbf{DG3}] \textbf{Create new classifiers in real-time:}
    The needs of the user can change dramatically over time and vary across users themselves. As an example, one user may wish to train a classifier to find data related to a specific hurricane event to expedite identification of people in desperate need of assistance. However, another user may wish to find data related to safety in general, not just a hurricane. As such, they should each be able to create and train their own classifiers in real-time specific to their needs at the time.
    
    \item[\textbf{DG4}] \textbf{Minimize model training time:}
    Although it is important to design a high-performing model, time constraints are equally important. Specifically, when the model is trained by user feedback, the user should not have to wait for several minutes for the model to be retrained and relabel data. Previously streamed data labels may update with retraining, allowing the user to potentially find important information that they had not seen before. As such, it is necessary to provide these updated results as quickly as possible for real-time situational awareness.
    
\end{enumerate}

\begin{figure}[!t]
    \centering
    \resizebox{\columnwidth}{!}{\includegraphics{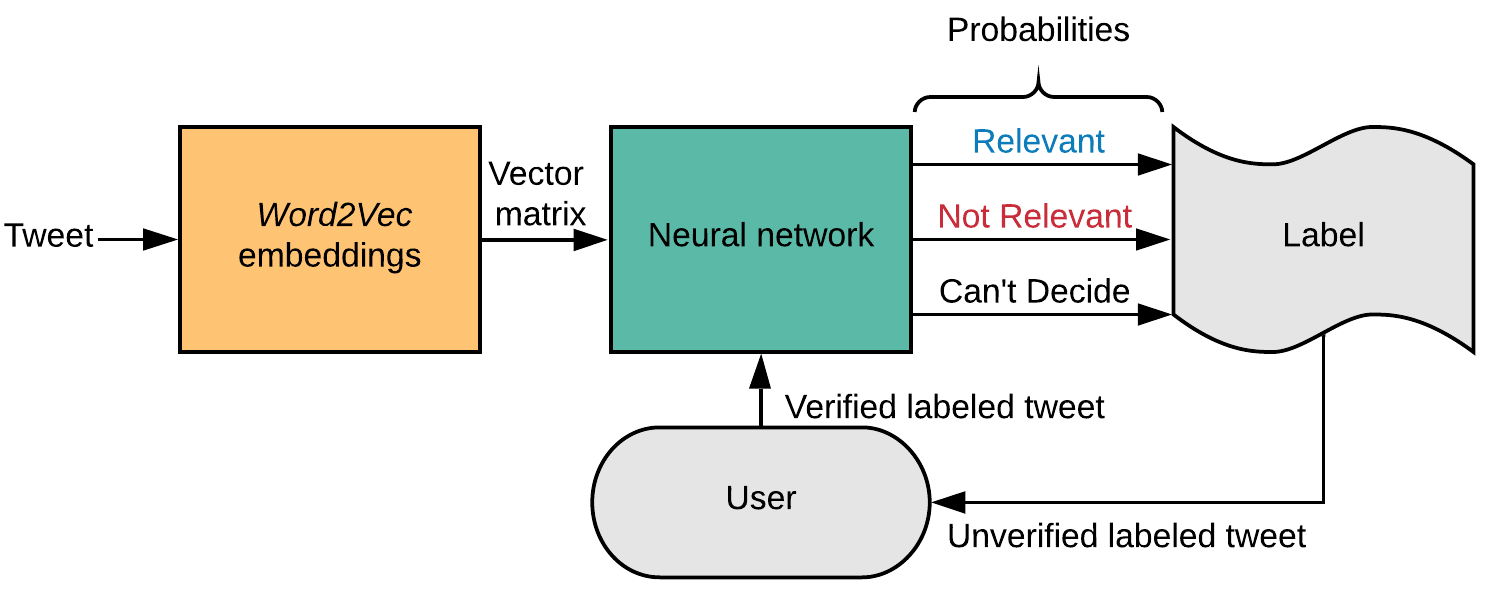}}
    \caption{High-level workflow of our framework with three main components: tweet vectorization, tweet classification, and user feedback.}
    \label{figure:workflow}
\end{figure}

\subsection{Workflow}
Fig.~\ref{figure:workflow} shows the three primary components of our framework's workflow applied to streaming tweets (however, the framework can be generalized to other kinds of text). First, as tweets are streamed in real-time, they are vectorized using a word embedding model.
Second, the vectorized tweets are provided as inputs to the neural network classifier (discussed in next section), which outputs a set of probabilities from the activation function of the tweet's predicted relevancy and assigns an unverified relevance label. Third, the labeled tweet is relayed to the user through the user interface. If the user identifies tweets with incorrect labels, they can  correct the label for the system to retrain and improve the model for relevance predictions. 

\subsection{Interactive Model Details}
In the following subsections, we elaborate on the underlying representations and models used to support our interactive learning framework. We design, optimize, and evaluate our approach with the key assumption that classifiers are trained (from scratch) in real-time using user-provided labels for streaming text. We simulate this process by adding training examples in small batches of 10 and evaluating against testing data, as explained below. All simulations were completed on a server with 128 GB RAM, 32 TB of disk storage, and 2 Intel(R) Xeon(R) E5-2640 v4 CPUs at 2.40GHz.

\subsubsection{Model Candidates}
Selecting the underlying model for our framework was a key task, as it must be efficiently trainable with a continual stream of user-labeled data (DG4). 
As discussed in Section 2, neural networks are a natural choice for online learning scenarios in which training data is supplied sequentially over time~\cite{NguyenDatTien2016AoOD, NguyenDatTien2016RCoC}. 
In addition, neural networks have generated impressive results with \texttt{Word2Vec}~\cite{MikolovTomas2013DRoW} embeddings~\cite{NguyenDatTien2016AoOD, NguyenDatTien2016RCoC, TangD.2014Lswe}.
Therefore, we employ a neural network as our classifier to determine text relevance based on real-time training examples provided by the user. To convert the text into vector inputs (of our neural network), we use word embeddings generated by Google's \texttt{Word2Vec} skip-gram model~\cite{word2vec, MikolovTomas2013DRoW}, which contains 3 million 300-dimensional word vectors pre-trained (and therefore, capturing word embeddings) on a subset of the Google News dataset with approximately 1 billion words.

In selecting the specific neural network model type, we experimented with the well-known Convolutional Neural Network (CNN)~\cite{lecun1998gradient}, Long-Short Term Memory (LSTM) Neural Network~\cite{hochreiter1997long}, and Recurrent Neural Network (RNN)~\cite{elman1990finding} since they have performed well in various text classification tasks~\cite{yin2017comparative}. Hybrid architectures, such as recurrent convolutional neural networks~\cite{LaiS.2015Rcnn}, have also been proposed in recent years, but have not been made available in well-supported libraries. Therefore,  we did not consider them in this paper, since our goal was to also support a well-tested SMART 2.0 system for end users. 

Our CNN model contains the traditional convolutional and max-pooling layers before activation~\cite{yin2017comparative}. Specifically, we apply a 1-dimensional convolutional layer, 1-dimensional max-pooling layer, flatten the output, and then activate it with softmax and a dense layer. The filter and kernel sizes of the convolutional layer are optimized during the hyperparameter stage (explained in Section 3.3.4). We use Hochreiter's LSTM ~\cite{hochreiter1997long} and the traditional RNN ~\cite{elman1990finding} architectures as provided by \texttt{Keras}~\cite{keras}. The LSTM and RNN hidden layer each contain 300 hidden neurons and use softmax activation.


\subsubsection{Design}
As mentioned before, to enable the use of neural networks for classifying text, we convert the unstructured text (of the tweets) into vectors ready for consumption by the neural network. 
When using \texttt{Word2Vec} vectors as features for classification, a common approach is to convert each word in the sentence to its vector, average the word vectors in the sentence, and then use the resulting feature vector for model training~\cite{abdelwahab2016uofl, TangD.2015Dmwg}. However, averaging the vectors results in the loss of syntactic information, which can negatively impact classification results~\cite{LingW.2015Tsao}.
As an example, the two sentences ``Only Mary will attend the ceremony." and ``Mary will only attend the ceremony." would generate identical averaged sentence vectors since they contain the same set of words, but they differ in meaning.
Therefore, to capture both semantic and syntactic information, we represent a sentence as a matrix where each row $i$ is a 300-dimensional \texttt{Word2Vec} vector corresponding to word $i$ in the original sentence. 

 The input to the neural network consists of the matrix representing the sentence (as described above) and the output consists of the classification labels for the input sentence (Fig.~\ref{figure:workflow}). Specifically, we allow a tweet to be (1) Relevant, (2) Not Relevant, or (3) Can't Decide. The label with the highest probability from the activation function corresponds to the final label given to it. The ``Can't Decide'' label indicates that the tweet may or may not be relevant depending on the context. This is useful if the user finds a social media post such as ``Remembering when Hurricane Irma destroyed my home...'' that may not directly relate to the current event, but may be semantically relevant, and the user does not want to mark such cases as ``Not Relevant''. This gives the user more flexibility to accommodate their needs since the definition of relevancy will depend on both the user and the situation. 


\subsubsection{Corpus for Model Selection and Optimization}
To experiment with different neural network model types and optimize the selected model, we used a disaster-related corpus annotated on the crowd-sourcing platform, Figure Eight~\cite{figureeight}. The dataset contains 10,876 tweets related to different types of disaster events, such as hurricanes and automobile accidents. The data was collected using keywords such as ``ablaze'' or ``quarantine'', and therefore, covers a wide variety of disaster-related topics. Our main motivation for using this open dataset is its size (as well as topical relevance), enabling the optimization of hyperparameters and comparison of various models. In the corpus, each tweet is manually labeled by Figure Eight's workers as ``Relevant", ``Not Relevant", or ``Can't Decide", and the distribution of labels is unbalanced. Specifically, there are 4,673 ``Relevant" instances, 6,187 ``Not Relevant" instances, and 16 ``Can't Decide" instances. This dataset has been used in other tweet classification research projects ~\cite{ToHien2017OIDT}. However, the researchers of that study remove the tweets with the ``Can't Decide" label to improve training data quality. As explained in the previous section, we find the ``Can't Decide" option useful for users to apply to cases with insufficient context for relevance determination.
We randomly shuffle the data and divide the dataset into 80\% training, 10\% validation, and 10\% testing sets.

It is important to note that we only use the Figure Eight dataset to optimize the hyperparameters and provide an initial evaluation of the model by simulating the provision of labels in real-time by the user. 
Since each tweet in the dataset contains true labels that were manually assigned by humans, it allows us to evaluate the model performance by comparing the model's predictions to the true labels after each training iteration.
Our proposed approach as well as its integration within the SMART 2.0 system, however, allows for the creation of the models from scratch (with no prior training) (DG3), leveraging real-time labels provided by users on streaming data for training.

\subsubsection{Optimization}
 \begin{table*}[!t]
\centering
\setlength\tabcolsep{0pt} 
\smallskip 
\begin{tabular*}{\linewidth}{@{\extracolsep{\fill}} ccccccccccccc}
\toprule
  Model & Learning& Batch & Epochs & Dropout & Recurrent & Filter & Kernel & Optimizer & Average & Average & Average & CPU \\
  &Rate&Size&&&Dropout&Size&Size&&Precision&Recall&$F_1$ score&Time (sec) \\
\toprule
  CNN& 0.0079 & 10    & 1   & -- & -- & 16 & 2 & Adam & 0.75 & 0.73 & \textbf{0.74} & \textbf{503.82} \\
  CNN& 0.01   & 50    & 2   & -- & -- & 16 & 2 & Adagrad & 0.73 & 0.71 & 0.72 & 522.47 \\
  CNN& 0.0063   & 10    & 3   & -- & -- & 16 & 2 & Adam & 0.73 & 0.71 & 0.72 & 553.43 \\
\midrule
  LSTM& 0.0002   & 10    & 10 & 0.4 & 0.2 & -- & -- & Adam & 0.7597 & 0.7475 & \textbf{0.7534} & 4241.97 \\
  LSTM& 0.0002   & 20    & 8   & 0.2 & 0.6 & -- & -- & Adam & 0.7597 & 0.7468 & 0.7530 & \textbf{4100.37}  \\
  LSTM& 0.0006   & 100    & 12   & 0.6 & 0.6 & -- & -- & Adam & 0.7559 & 0.7431 & 0.7493 & 4209.37 \\
\midrule
  RNN& 0.0001   & 10    & 7   & 0.0 & 0.2 & -- & -- & Adam & 0.7037 & 0.6957 & \textbf{0.6996} & 3069.81 \\
  RNN& 0.0001   & 20    & 5   & 0.0 & 0.0 & -- & -- & Adam & 0.7028 & 0.6921 & 0.6973 & \textbf{2805.52}  \\
  RNN& 0.0001   & 100    & 12   & 0.0 & 0.2 & -- & -- & Adam & 0.70 & 0.69 & 0.69 & 3160.35 \\
\bottomrule
\end{tabular*}
\vspace*{3mm}
\caption{
Average precision, recall, $F_1$ score, and CPU time for the top three performing hyperparameter combinations on each of the CNN, LSTM, and RNN models. Bold numbers correspond to the highest $F_1$ scores and lowest CPU times for each of the three model types. We report the recall, precision, and $F_1$ score to four decimal places (when necessary) to distinguish the average $F_1$ scores. }
\label{table:hyperparameters}
\end{table*}

\begin{table}[!h]
\centering
\smallskip 
\begin{tabular*}{1\columnwidth}{@{\extracolsep{\fill}} ccccc}
\toprule
  Model & Average & Average & Average & CPU \\
  &Precision&Recall&$F_1$ score&Time (sec) \\
\toprule
  CNN& 0.74 & 0.73 & 0.73 & \textbf{501.10} \\
  LSTM& 0.76 & 0.74 & \textbf{0.75} & 4211.01 \\
  RNN& 0.70 & 0.69 & 0.70 & 3085.59 \\
\bottomrule
\end{tabular*}
\vspace*{3mm}
\caption{
Testing results with the optimal hyperparameter combinations for the CNN, LSTM, and RNN models. The bold numbers correspond to the highest $F_1$ score and lowest CPU time among the three models.}
\label{table:testingResults}
\end{table}

In order to experiment with the different neural network model types, we ran several training simulations with random combinations of hyperparameters (i.e., random grid search) to see which model converged to the best $F_1$ score.
The $F_1$ score is a metric widely used to evaluate the quality and performance of machine learning models and neural networks~\cite{SokolovaM.2006BaFa}. It is computed as the harmonic mean of \textit{precision} (the proportion of true positive predictions compared to the total number of positive predictions) and \textit{recall} (the proportion of true positive predictions compared to the overall number of positive instances) : $F_1 = \frac{2 \times precision \times recall}{precision + recall}$.
\textcolor{red}{The $F_1$ score provides a balanced measure, combining these two performance aspects. It is therefore more informative compared to other metrics such as accuracy, especially when the training and testing sets are imbalanced~\cite{dal2015calibrating}, as in our case}.

A central part of our approach to the training, validation, and verification of learning models is simulating the interactivity of visual analytics for real-time data, i.e. for use cases in which training data does not exist prior to user interaction. We assume the user (re)labels the incoming stream of data and therefore iteratively trains a model, which consequently meets their real-time needs.
To replicate this process, we computationally evaluate the model's performance (as if it is successively trained by user-labeled data) by iteratively training the model with 10 new samples from the training dataset.
We average the $F_1$ score obtained from each of these iterations and use the resulting number to measure the model's performance.
In addition, we introduce a new variable, \texttt{window size}, for our training iterations. 
Specifically, due to the considerably small amount of training data provided by the user, we found that an appropriately small number of epochs (one forward and one backward pass over the training data in the model) was necessary to reduce performance degradation from initial overfitting.
However, we also found that increasing the number of epochs could lead to higher $F_1$ scores as more data was provided.
Thus, we use a sliding window of 110 samples that includes the (successively provided) new training data (10 samples) as well as the most recently used training data (100 samples) to both account for small amounts of training samples and increase the number of total training epochs for a given sample.


We use the validation data to optimize the hyperparameters for each of the CNN, LSTM, and RNN models. Specifically, after each training iteration with 10 new samples, we evaluate the neural network's $F_1$ score on the validation set to view its simulated performance as if it was trained by gradual user labeling. After identifying the optimal hyperparameters for each of the CNN, LSTM, and RNN models, we evaluate their performance on the testing set.

Table~\ref{table:hyperparameters} demonstrates the results from our validation stage. Specifically, it lists the average $F_1$ score obtained during each training simulation along with the total CPU time required to complete the simulation \textcolor{red}{(accumulated with each training and evaluation iteration)}.
Although in many applications, $F_1$ score alone is sufficient to evaluate machine learning models, it is not for ours. 
To see why, note that the LSTM model yields an $F_1$ score of 0.75, the highest of any hyperparameter combination.
However, the LSTM model (with the highest $F_1$ score) takes approximately 4,242 seconds to complete training, whereas the CNN model (with the highest $F_1$ score) only takes 504 seconds.
Thus, the LSTM model takes roughly eight times longer to simulate than the CNN model, but does not improve its $F_1$ score by a significant amount (LSTM: 0.75 vs. CNN: 0.74).
In the context of interactive learning, we wish to balance the training/CPU time and performance
such that the model both performs well and retrains in a short amount of time for rapid improvement (DG4). 
Therefore, it is necessary to consider both the CPU time and average $F_1$ score.
With these optimization standards in mind, we chose the hyperparameters that yielded the highest $F_1$ scores for each model since the other hyperparameter combinations generated lower $F_1$ scores and higher or comparable CPU times.
The selected combinations correspond to rows 1, 4, and 7 in Table~\ref{table:hyperparameters} with the average $F_1$ scores in bold.

\begin{figure}[!t]
    \centering
    \resizebox{0.95\columnwidth}{!}{\includegraphics{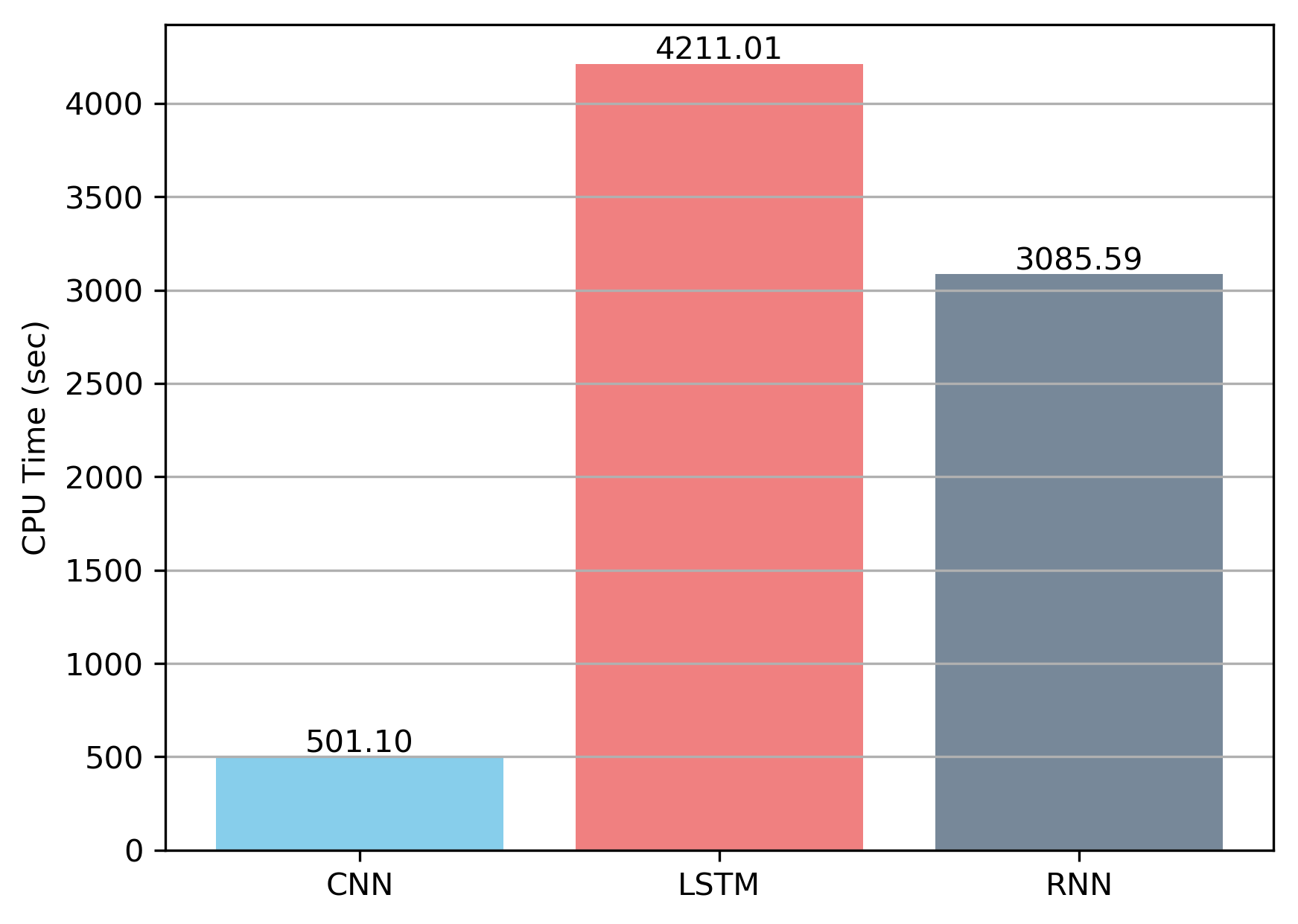}}
    \caption{The total CPU time required for each model to complete the testing simulation. The CNN model is noticeably faster than both the LSTM and RNN models. }
    \label{gpu_time}
    \vspace{-3mm}
\end{figure}

The testing process is identical to the validation process: after the model is trained with 10 new samples, its performance is measured by computing the average $F_1$ score on the testing set (using the optimized hyperparameters from the validation stage). 
Our results are summarized in Table~\ref{table:testingResults}.
We found that the LSTM model yielded the highest $F_1$ score of 0.75. The CNN and RNN models achieved a 0.73 and 0.70 $F_1$ score, respectively. 
Based on these results and the previously discussed optimization standards, we selected the optimized CNN model for our classifier. 
In particular, the CNN simulation not only yielded a competitive average $F_1$ score of 0.73, but also achieved this score 6 to 8 times more quickly than the LSTM or RNN (Fig.~\ref{gpu_time}), which is significant in terms of responding to user feedback in a timely manner.

\textcolor{red}{The optimized CNN model yielded 0.74 and 0.73 average precision and recall scores respectively (Table~\ref{table:testingResults}, row 1).
This model performance may be due to the initial lack of sufficient training data and difficulty in classifying certain tweets. For instance, after examining the testing dataset, we found that many misclassified tweets were extremely short (e.g., the tweet ``screams internally'' was misclassified as ``Relevant'') or contained complex disaster-related diction (e.g., the tweet ``emergency dispatchers in boone county in the hot seat'' was misclassified as ``Relevant''). However, as we demonstrate in the next section, our model still outperforms state-of-the-art learning models on tweet datasets.} 

It is worth noting that we do not save the trained model from the validation or testing stages for evaluation in the next stage (or for use with SMART 2.0). We only save the optimized hyperparameters. 
This is because we assume that users start training a new model (for any event or topic they choose) by labeling the incoming stream of tweets. 
  
In this section, we optimized the model on a sufficiently large dataset that contained tweets related to several kinds of disasters. In the next section, we evaluate the model on datasets containing tweets on specific events, which is representative of cases for situational awareness.

\subsubsection{Evaluation}

To further demonstrate the optimized CNN model's performance, we computationally evaluated it on wildfire, bombing, and train crash datasets from CrisisLexT26~\cite{OlteanuAlexandra2015WtEW}, each of which contain approximately 1,000 tweets collected during 2012 and 2013 from 26 major crisis situations labeled by relevance. 
We apply a similar process to evaluate our optimized CNN model on these datasets as we did with the Figure Eight~\cite{figureeight} dataset.
Specifically, we split the data into 50\% training and 50\% testing sets \textcolor{red}{(to replicate the experimental setting of To et al.~\cite{ToHien2017OIDT}, against which we will compare our results)}, train the model by supplying 10 tweets from the training set at a time (to simulate user labeling of streaming data), evaluate the resulting model on the entire testing set, and then average the $F_1$ scores for each evaluation.

\begin{figure}[!t]
    \centering
    \resizebox{0.84\columnwidth}{!}{\includegraphics{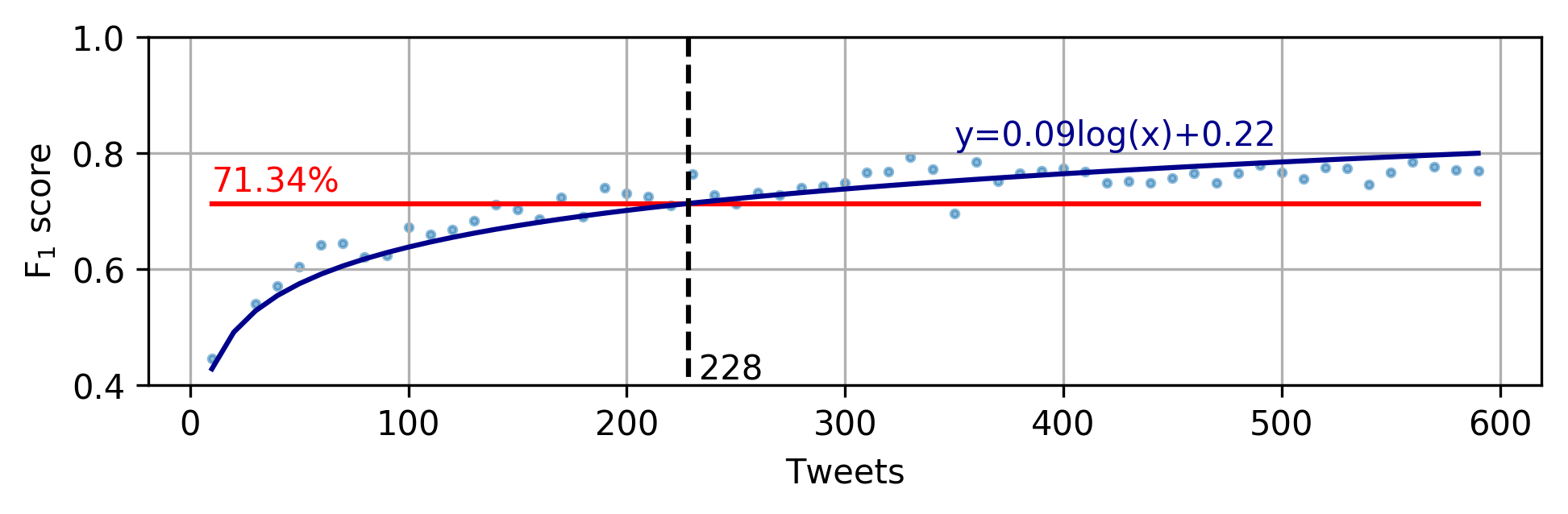}}
    \caption{Optimized CNN $F_1$ score per training iteration of 10 tweets with the Colorado wildfires dataset (Table~\ref{modelEval}). The $F_1$ scores are logarithmically fitted and intersect with the average $F_1$ score (0.7134) at 228 tweets.}
    \label{wildfires}
\end{figure}

\begin{figure}[!t]
    \centering
    \resizebox{0.84\columnwidth}{!}{\includegraphics{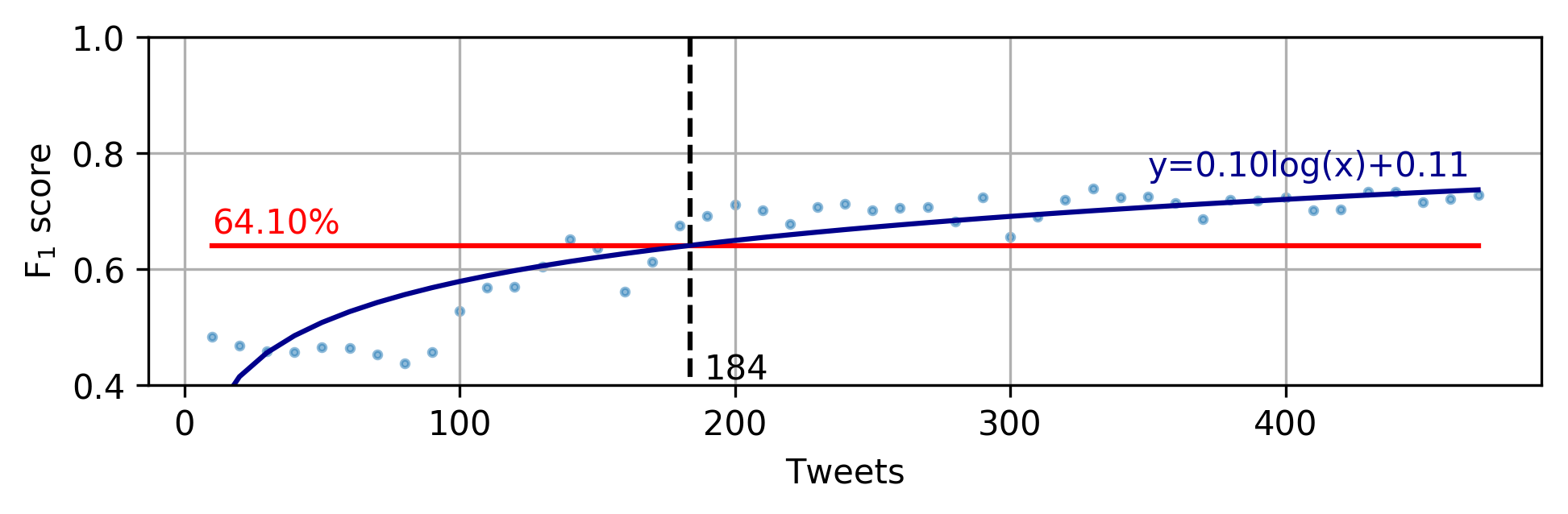}}
    \caption{Optimized CNN $F_1$ score per training iteration of 10 tweets with the Boston bombings dataset (Table~\ref{modelEval}). The $F_1$ scores are logarithmically fitted and intersect with the average $F_1$ score (0.6410) at 184 tweets. }
    \label{bombings}
\end{figure}

\begin{figure}[!t]
    \centering
    \resizebox{0.84\columnwidth}{!}{\includegraphics{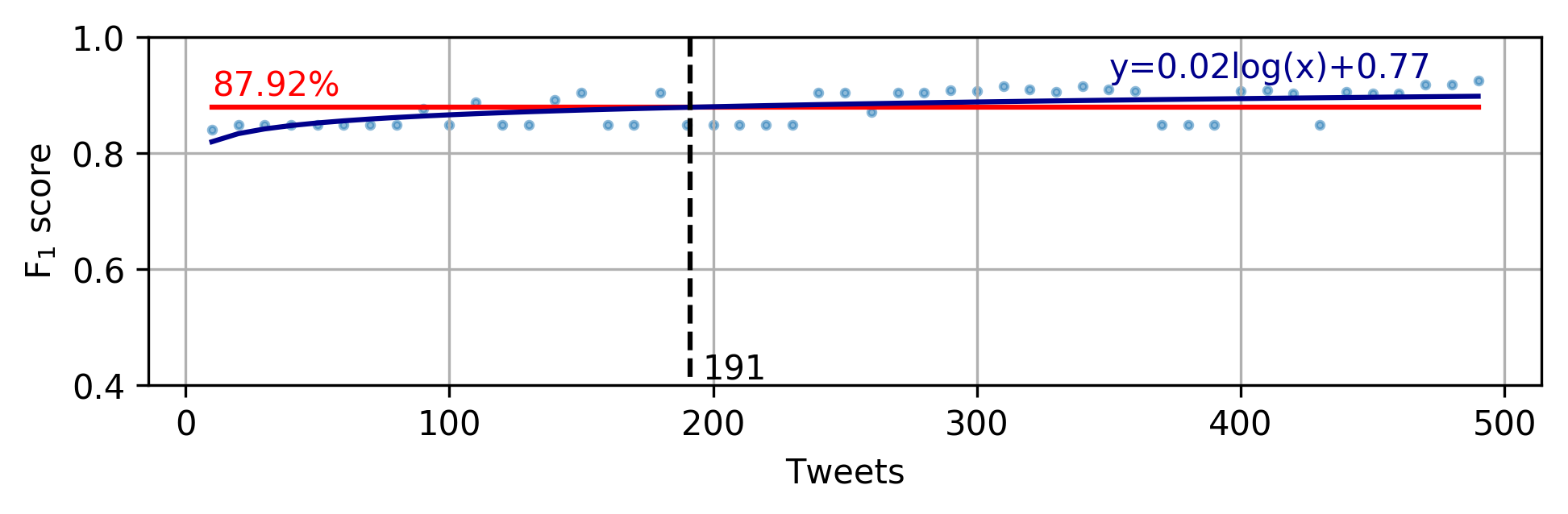}}
    \caption{Optimized CNN $F_1$ score per training iteration of 10 tweets with the NY train crash dataset (Table~\ref{modelEval}). The $F_1$ scores are logarithmically fitted and intersect with the average $F_1$ score (0.8792) at 191 tweets.}
    \label{traincrash}
\end{figure}

We summarize our results in Table~\ref{modelEval} and graph the model's performance for retraining with 10 new incoming tweets in Fig.~\ref{wildfires},~\ref{bombings}, and~\ref{traincrash}. \textcolor{red}{In addition, we report the average CPU times to train the model during a single iteration (10 tweets) with each dataset in Table~\ref{modelEval}. Since the datasets vary slightly in size, we only compute the averages from the first 45 iterations since the smallest dataset (Boston Bombings) required 45 iterations to complete the simulation. We found that per-iteration training was fast and approximately 0.5 seconds with each dataset, which meets our timing demands (DG4). }

We obtained 0.71, 0.64, and 0.88 $F_1$ scores for the Colorado wildfires, Boston bombings, and NY train crash datasets, respectively. Interestingly, the variance of the $F_1$ scores over the datasets is significant. 
The textual data in the Boston bombings dataset, which yielded the lowest average $F_1$ score, was not as easy to separate into the different relevance categories by the model compared with the other two datasets.
However, the $F_1$ score does eventually converge towards a higher value similar to the other datasets, indicating the potential presence of outliers during the first few training iterations.
In addition, we found that the simulations converged to the average $F_1$ scores after training with approximately 190--230 tweets, depending on the dataset, meaning that users need to label 190--230 tweets to achieve the reported $F_1$ scores. 
However, the CrisisLexT26 datasets also correspond to specific events, such as wildfires.
As such, we surmise that interactively training the model on specific, well-defined events will reduce the amount of training data needed to achieve satisfactory results than with generic constraints on relevance (e.g., a classifier about safety in general).

\begin{table}[!t]
\centering
\smallskip 
\begin{tabular*}{1\columnwidth}{@{\extracolsep{\fill}} ccccc}
\toprule
  Category & Average & Average & Average & Average CPU \\ 
  &Precision&Recall&$F_1$ score & Time (sec) \\ 
\midrule
  Colorado wildfires & 0.72 & 0.71 & 0.71 & 0.49 \\
  Boston bombings & 0.64 & 0.65 & 0.64 & 0.50 \\ 
  NY train crash & 0.86 & 0.90 & 0.88 & 0.49\\ 
\bottomrule
\end{tabular*}
\vspace*{3mm}
\caption{
Average precision, recall, and $F_1$ score for three CrisisLexT26~\cite{OlteanuAlexandra2015WtEW} datasets.  }
\label{modelEval}
\end{table}


Finally, we compare our results with the learning-based algorithm employed by To et al.~\cite{ToHien2017OIDT}, who also evaluated their model's performance with CrisisLexT26 datasets. In particular, their learning-based approach used \texttt{Word2Vec}, TF-IDF, latent semantic indexing, and logistic regression for classifying data as relevant or irrelevant. The authors of that study split the dataset into two equal parts: one for training and one for testing. They trained the model once (as opposed to our iterative approach) and evaluated on the testing set. Their algorithm was able to yield high precision scores between 0.85--0.95, compared to our scores of 0.64--0.86. However, their recall scores were approximately 0.22--0.45, considerably lower than our recall scores of 0.65--0.90. Therefore, our approach outperforms the learning-based model presented by~\cite{ToHien2017OIDT}, in terms of the overall $F_1$ score: our interactive approach achieves $F_1$ scores of $0.64$--$0.88$ (depending on the dataset) compared to $0.45$--$0.64$ by~\cite{ToHien2017OIDT}. The authors also presented a matching-based approach that achieved a much higher $F_1$ score of 0.54--0.92, which is comparable to ours. However, they generate the set of hashtags to be used for matching by scanning all of the tweets in the dataset. Since we assume the data is streamed in real-time, and therefore, not available altogether, we use an iterative learning approach.

\section{SMART 2.0}

\subsection{SMART}
\begin{figure*}[!t]
    \centering
    \includegraphics[width=\linewidth]{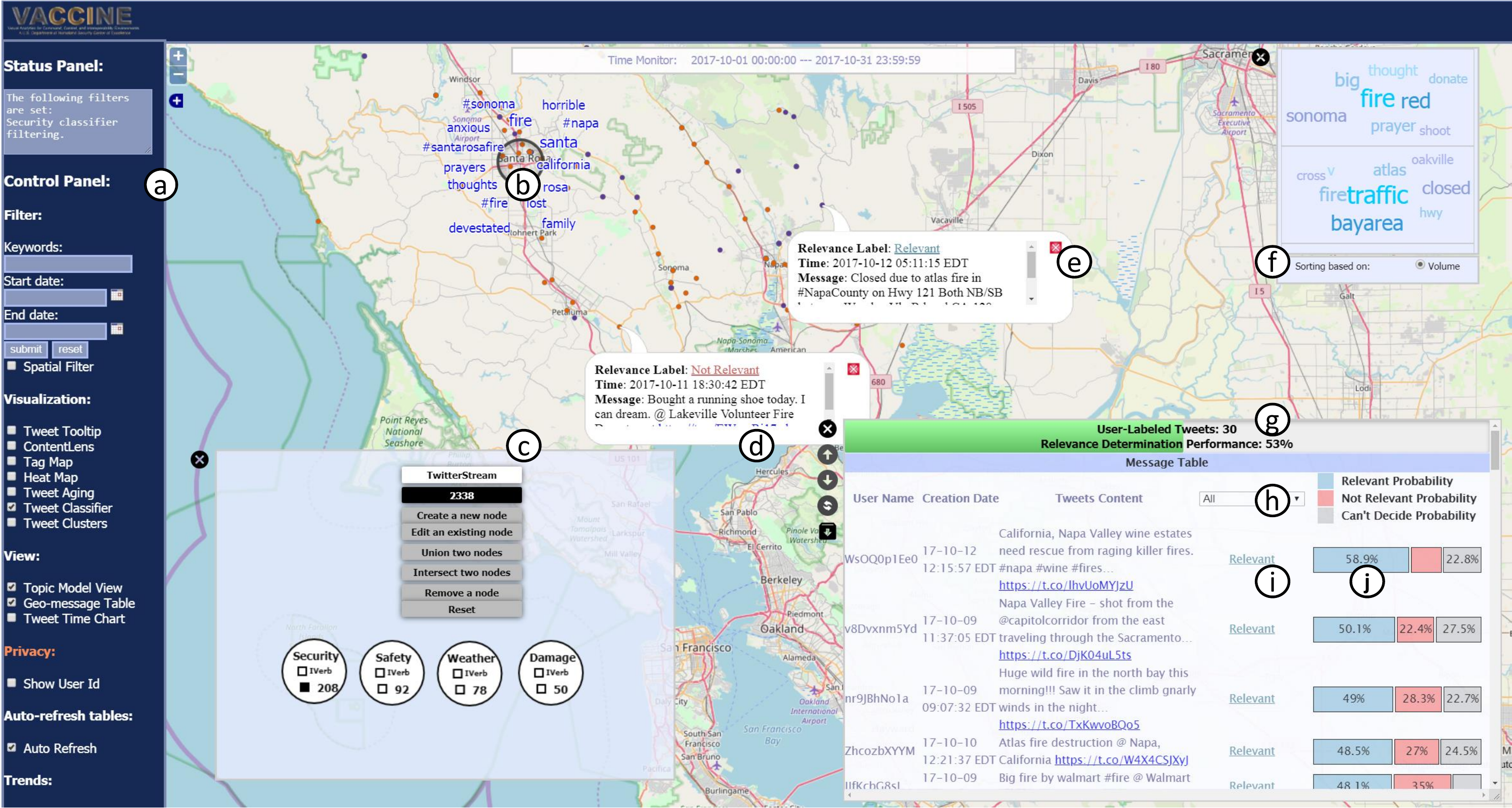}
    \caption{
    SMART 2.0 overview: (a) The control panel provides several filters, visualizations, and views. (b) The content lens visualization provides the most frequently used words within a selected area. (c) The tweet classifier visualization provides keyword-based filters to help reduce noisy data. (d)(e) Clicking a tweet on the map with the tweet tooltip visualization displays the tweet's time, message, and relevance label. (f) The topic-modeling view, based on Latent Dirichlet Allocation, extracts trending topics and the most frequently used words associated with each topic among tweets with specified relevancy. (g)--(j) The message table aggregates the tweets for efficient exploration with (g) \textcolor{red}{the model's estimated classification performance ($F_1$ score)}, (h) a drop down box to filter data by their relevance labels, (i) color-coded relevance labels that can be changed by clicking on the label itself, and (j) associated relevance probabilities. \textcolor{red}{Tweet map symbols are colored orange and purple to distinguish Twitter data from Instagram-linked tweets, respectively, since the latter contains potentially useful images for situational awareness.} }
    \label{fig:smart}
\end{figure*}

The Social Media Analytics and Reporting Toolkit (SMART)~\cite{zhang2017smart, ZhangJiawei2014Riam} 
is a visual analytics application designed to support real-time situational awareness for first responders, journalists, government officials, and special interest groups.
SMART obtains real-time publicly available geo-tagged data from the Twitter streaming API. 
The user is able to explore the trending and abnormal topics on various integrated visualizations, including spatial topic model visualization and temporal views.
The tweet time chart and theme river visuals convey the temporal distributions of topics if the user wishes to determine how the content of streamed social data has changed over time.

SMART uses string matching-based classifiers to visualize relevant data. 
Specifically, the user can either (a) select pre-defined filters, such as \textit{Safety} or \textit{Weather} (Fig.~\ref{fig:smart}(c)), each using a series of related keywords for inclusion and exclusion of tweets in the subsequent topic-modeling (Fig.~\ref{fig:smart}(f)) and (geo)visualizations (Fig.~\ref{fig:smart}(b)), or (b) create their own filters by supplying keywords, and intersect or union multiple filters according to their needs.
However, keyword-based matching is insufficient for finding relevant information as it fails to accurately capture semantic relevance and therefore effectively filter out noisy data.
As an example, if the user were to apply the \textit{Safety} classifier, it would be possible for the tweets ``My house is on fire!'' and ``I was just fired from my job.'' to pass through the filter since they both include the keyword \textit{fire}. However, the latter is unrelated to the intended semantic context of \textit{Safety} and thus dilutes the filter's quality.

To address this problem, we integrate our interactive learning framework (the focus of this paper) in the existing SMART application~\cite{zhang2017smart, ZhangJiawei2014Riam} and seek domain expert feedback on the use of these models.
We call the resulting extended application SMART 2.0. SMART 2.0 allows users to define string matching-based keyword filters (similar to SMART), but adds the ability for users to then iteratively refine and train the newly integrated models by labeling the filtered data as semantically relevant or not.
In addition, the SMART 2.0 interface includes interactive visuals to facilitate user exploration, filtering, and refinement of relevant data (Fig.~\ref{fig:smart}).

\textcolor{red}{As with the model simulations in Section 3.3, SMART 2.0's underlying models are trained with successive batches of 10 user-labeled tweets.}
\textcolor{red}{In cases where model predictions conflict with user labels, user labels override the model's since they represent the ground truth.
In addition, users should not need to manually relabel the same data multiple times.
Although conflicts might indicate that the model is not sufficiently trained, 
the model trains with the same data during several successive iterations (as discussed in Section 3.3.4),
so conflicts might be resolved after future iterations.
}

\subsection{SMART 2.0 Interface}
\textcolor{red}{The extensions to SMART 2.0's user interface, compared with SMART, concern the new interactive visuals that allow users to iteratively train machine learning models, utilize model predictions for rapid relevancy identification, and  understand a model's reliability.}
The SMART 2.0 interface (Fig.~\ref{fig:smart}) extends the interactive features of SMART for relevance identification in three primary ways:
\begin{enumerate}
    \item Extending the tweet table (containing a tweet's creation date and text) by including the predicted relevance label, relevance label probabilities, label modification, \textcolor{red}{model training performance, and relevance filtering}.
    \item Extending the interactive map containing the geo-tagged tweets whose relevancy can be individually inspected or modified.
    \item Altering the content of existing SMART views (e.g., topic models and spatial topic lenses) using either all data or only relevant data (as identified by the model and corrected by the user).
\end{enumerate}

\subsubsection{Table}
    The SMART 2.0 table (Fig.~\ref{fig:smart}(g)--(j)) is extended from SMART in that it not only provides a tweet's creation date and text, but also provides the predicted relevance label (Fig.~\ref{fig:smart}(i)) and the probabilities of a tweet belonging to any of the relevance classes (Fig.~\ref{fig:smart}(j)) (DG1). 
    
    In particular, the relevance of a tweet can be ``Relevant", ``Not Relevant'', or ``Can't Decide". The ``Relevant'' label is colored blue, the ``Not Relevant" label red, and the ``Can't Decide" label gray to visually separate tweets with different relevance.
    \textcolor{red}{SMART's preexisting blue color scheme motivated us to use the blue, red, and gray diverging coloring for relevancy in order to maintain visual appeal and harmony. 
    }
    
    Users can directly click on relevance labels to correct the classifier's prediction (DG2).
    For instance, if a tweet is incorrectly marked ``Relevant", clicking the label will change it to ``Not Relevant" or ``Can't Decide", depending on the label the user wishes to assign. 
    Further, a drop down box is included at the top of the relevance label column (Fig.~\ref{fig:smart}(h)), which provides the option to filter out data that does not have a specified relevancy (DG1).
    For example, by selecting ``Relevant" from the drop down box, the table will remove tweets with labels ``Not Relevant" and ``Can't Decide" from all views and visualizations in SMART, including geovisualizations and temporal views.
    
    The table also displays the \textit{degree} (or confidence) of a tweet's relevancy.
    In specific, the probabilities of a tweet being ``Relevant", ``Not Relevant", or ``Can't Decide" are represented as a horizontal segmented bar graph and sized proportional to their respective percentages (Fig.~\ref{fig:smart}(j)). 
    In addition, the user can sort tweets based on relevancy probability in ascending or descending order.
    
    We provide the relevance probabilities and associated sorting actions as a supplementary relevance filtering mechanism (DG1). In particular, it is possible for tweets to be classified as ``Relevant" by the model, for example, but with low confidence. The probability filtering allows the user to specifically view high-confidence relevant data and therefore further reduce potentially noisy data.
    
    \textcolor{red}{The table provides a performance bar that encodes the estimated performance ($F_1$ score) of the underlying learning model (Fig.~\ref{fig:smart}(g)), as well as the number of user-labeled tweets, to inform the user of the model reliability. Since labeled testing data is not available to evaluate the model for real-time training (because we assume the user may train on any type of event data and has their own specifications for relevancy), the model's performance can only be estimated.
    Based on our evaluations in Section 3.3.5 with datasets typical of situational awareness scenarios (Table~\ref{modelEval}), the Colorado wildfires dataset generated the $F_1$ score (0.71) closest to the average of the three datasets (0.74). Therefore, we use the Colorado wildfires dataset's logarithmic trendline $y = 0.09\log_e(x) + 0.22$ (Fig.~\ref{wildfires}) to approximate the model's $F_1$ score as a function of the number of user-labeled tweets.
    }
    
\subsubsection{Map}
    The SMART 2.0 map is extended from SMART in that it includes a tweet's relevance label (which can be modified) in addition to its text and creation date (Fig.~\ref{fig:smart}(d)(e)).
    Through the \textit{Tweet Tooltip}, the user can directly click on tweet symbols on the map to view their text and associated relevancy (DG1).
    In addition, the user can correct the classified relevance label (DG2) by clicking on the label itself. 
    Map inspection can allow the user to view and investigate potential geographical relevancy trends. For example, during crisis events, relevant tweets might be closely grouped on the map, so it may be more beneficial for the user to view predicted relevance from the map itself. 

    The interactions between the table and map are synchronized. If the user relabels data on the map, the associated new label will also be updated in the table, and vice versa. In addition, selecting a relevancy filter from the drop down box in the table filters the tweets on the map. 
    
    
\subsubsection{Integration with Existing Visualizations}
    Many of SMART's original visualizations, such as the topic-model views, spatial topic lenses, and temporal views help users make sense of spatiotemporal text data. Therefore, we integrated all of these views in SMART 2.0 with the relevance extensions.
    
    Users have the option to view only relevant or all the data (including irrelevant tweets) in various visualizations in case the interactive classifiers are not yet trained to desirable accuracies since, as we show in Section 3.3.5, classifiers typically require around 200 user-labeled tweets to achieve $F_1$ scores of 0.70--0.80. 
    If they choose to view only relevant tweets, any relevance filtering action also updates the data used by other visuals. 
    For example, the topic-modeling view (Fig.~\ref{fig:smart}(f)) extracts the top 10 topics from the tweets and displays the most frequently used words for each topic.
    If the user filters out irrelevant tweets, the topic-modeling view will only be applied to the remaining relevant tweets. 
    \textcolor{red}{It is important to note that the majority of visualizations in SMART 2.0  require a minimum number of tweets in order to render.
    When filtered relevant data is scarce, visualizations do not populate, in which case users can individually inspect tweets.
    For instance, the topic-modeling view requires at least 10 tweets to extract topics.}
    
    Overall, SMART 2.0's suite of visualization tools can be used in combination with relevance interactions to further understand trends and important spatiotemporal characteristics of relevant data.
    


\section{User Experience}
In this section, we provide usage scenarios and feedback from domain experts that demonstrate our framework's effectiveness and usability.

\subsection{Usage Scenario 1} 
Alice is an emergency dispatcher interested in identifying people in need for help or hazardous locations during a hurricane. She uses SMART 2.0 to find any related social media posts near the affected area. She adds a new filter \textit{Hurricane} and provides an appropriate set of filter keywords such as ``hurricane", ``help", ''blocked'', and ``trapped". 

After applying the \textit{Hurricane} filter, she explores the filtered tweets in the table and finds a tweet labeled ``Relevant" that says ``Does anyone know how to get help setting up my TV?". Since the tweet is unrelated to a hurricane, she relabels it as ``Not Relevant''. After further browsing the table, she finds a tweet that says ``The road near Taylor Loop is blocked from a broken tree.", but it is labeled as ``Not Relevant''. Since the tweet contains actionable information, she relabels it as ``Relevant". After labeling several more tweets for model training and noticing that the model predicts correctly, she decides to only view ``Relevant'' tweets and sort them by most relevant. She promptly identifies a tweet posted only a few minutes ago marked as highly relevant. It reads ``Car just crashed into tree blocking road near Taylor Loop!''. Alice immediately notifies first responders of the location to provide assistance.

By using SMART 2.0, Alice is able to identify important, relevant data more quickly through interactively training the model to remove noise and then filtering by relevance.

\subsection{Usage Scenario 2}
\begin{figure*}[!t]
    \centering
    \fbox{\includegraphics[width=0.99\textwidth]{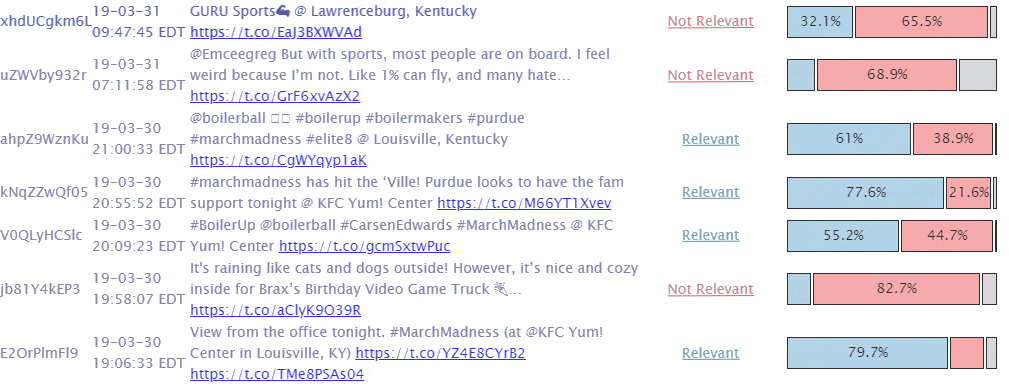}}
    \caption{Tweets with correctly predicted relevancy from the Purdue vs. Virginia 2019 March Madness game after the user (re)labels 80 tweets.}
    \label{fig:bbUsage}
\end{figure*}

To demonstrate the generalizability of our framework to other domains, we applied our interactive framework in real-time during the Purdue vs. Virginia 2019 March Madness game in the Kentucky area. 
We assumed the role of a journalist who wanted to follow public discourse on the game by identifying the relevant tweets.
We first constructed a \textit{Sports} filter, which included keywords such as ``Purdue'', ``game'', ``score'', and ``\#MarchMadness''. 
We then interacted with the streaming data by iteratively labeling the relevancy of tweets (from scratch) and found that the system correctly classified incoming data after roughly 80 training samples (Fig.~\ref{fig:bbUsage}). 
We noticed that the time intervals between successive trainings increased, indicating that it was more difficult to find incorrectly labeled data towards the end and that the model gradually learned from user feedback. In particular, the interval between the first and second training iterations was 2 minutes, whereas the interval between the final two was 4 minutes.

\subsection{Domain Expert Feedback}
We piloted SMART 2.0 with two groups of first responders, \textcolor{red}{each containing two individuals,} who frequently use SMART during events for situational awareness in their operations. 
Both groups participated in separate 1-hour long sessions \textcolor{red}{via conference call} in which they iteratively trained a classifier from scratch and applied relevance filtering and visualizations to assess the implemented framework.
\textcolor{red}{They received a tutorial of SMART 2.0 30 minutes beforehand and were provided with web access to the system to complete the session.}
For both groups, we simulated the real-time use of SMART 2.0 by feeding in a stream of historical data on events (previously collected).
For the first group, the system presented unlabeled tweets from the Las Vegas shooting on October 1, 2017 in the Las Vegas area. For the second group, we used unlabeled tweets from the October 2017 Northern California wildfires.
We used historical event datasets to ensure the existence of sufficient training relevant samples for a situational awareness scenario.

The domain experts in the first group applied the \textit{Safety}, \textit{Damage}, and \textit{Security} filters during the iterative training process, resulting in 317 tweets.
They trained on the same underlying model for all three filters, as they considered them semantically related.
In total, after relabeling approximately 200 tweets, they indicated that they could trust the model to predict accurately and were pleased that the tweets they had not seen before from the \textit{Security} filter were labeled correctly.
Their definition for relevancy was tweets containing actionable information.
For instance, they marked tweets containing information about road closures, blood drive locations, or death counts as relevant. 
They labeled data with general comments regarding the shooting, such as ``I hope everyone is safe now...terrible shooting...'', as irrelevant since they did not provide actionable information.
Interestingly, they also marked tweets that would influence public opinion (and therefore may cause action) such as those from bots or trolls as relevant since they still contained actionable information.

The domain experts from the second group followed a similar process in which they applied the \textit{Safety}, \textit{Damage}, and \textit{Security} filters, resulting in 445 tweets, and trained a learning model for relevance. 
They found that after training on roughly 67 tweets, the model satisfactorily predicted relevancy. 
As with the first group, these domain experts labeled tweets as relevant if they contained actionable information.

The domain experts from both groups \textcolor{red}{found SMART 2.0 to be easy to use and effective in identifying important data}. 
For instance, they discovered relevant, actionable information after training the model: specific blood drive locations to aid shooting victims.
Notably, the users mentioned that they felt less the need to relabel data as they progressed since the system provided more correct labels.
They were pleased that they had the option to view only relevant data, but could see all of the data regardless of relevancy
to avoid potentially missing important misclassified data, and that the model was responsive to user training.
In addition, they found the relevance percentage bars to be helpful in determining the tweets that were potentially the most relevant. 

One concern the domain experts had was that SMART 2.0 does not indicate the number of tweets that are predicted as relevant. 
They felt this extension could help them infer the occurrence of events or potential crises. 
For example, the number of relevant tweets for the \textit{Safety} classifier would likely increase significantly during a widespread disaster. 
\textcolor{red}{We plan to introduce this feature in the next development cycle.
However, we have added a visualization of estimated model performance in SMART 2.0 (Fig.~\ref{fig:smart}(g)) to help users ascertain the reliability of a model's relevancy predictions.}

Overall, the feedback from the domain experts was positive and helpful, indicating the system's practicality and usefulness in facilitating real-time situational awareness. 
In addition, they have asked to use SMART 2.0 in their emergency operations center.


\section{Discussion and Future Work}
Our interactive learning framework and SMART 2.0 integration were developed with the user in mind, influencing all of our design, computational evaluation, and implementation choices. 
Our user-centered model and SMART 2.0 application contribute to both the machine learning and visual analytics communities.
We bridge the two fields by demonstrating how models can be interactively trained and evaluated while keeping the user in mind, and used to facilitate situational awareness for real-life, practical use.

\textcolor{red}{SMART 2.0 currently only collects English tweets, although
supporting non-English languages (one at a time) with our current design (e.g., Spanish only) is straightforward since \texttt{Word2Vec} embeddings can be independently trained on a corpus in the target language~\cite{berard2016multivec}.
Extending our system to support multilingual tweets would be a powerful asset, especially for multilingual users, in amplifying situational awareness by leveraging relevancy of tweets issued in different languages. 
However, the multilingual model performance evaluation and testing is an open area for research.
In addition, 
determining the specifics of how a single relevance classifier might be trained with multilingual tweets requires careful attention. For instance, training iterations with Spanish tweets should also affect the relevancy of semantically-related English tweets. Since similar words in different languages likely have different vector representations (embeddings), multilingual mappings must be learned or training must be performed differently, such as with parallel corpora ~\cite{berard2016multivec}. Multilingual support also requires changes in SMART 2.0's language-dependent visualizations, such as the topic-modeling view (Fig.~\ref{fig:smart}(f)). Translation to a unified language or extracting topics separately for each language are two potential solutions. 
}

The scalability of our framework is a natural concern, especially since SMART and many deployed real-time visual analytics applications contain multiple users who require responsive interfaces while monitoring crisis events. We deliberately designed the framework architecture with scalability in mind. As mentioned in Section 3.3.4, we selected the model and optimal hyperparameters based on training/CPU time in an effort to maximize the model's computational speed. Further, SMART 2.0 filters and views \textit{at most} 800-900 tweets at a time, \textcolor{red}{although user-specified filtering (typical in situational awareness scenarios) reduces the data to only a few hundred tweets, as demonstrated in Section 5.3.} It takes only 2-3 seconds to calculate and retrieve their relevance labels over the network from the server where the model resides, \textcolor{red}{and per-iteration training is fast, as established in Section 3.3.5.}


Training a model during a particular event, such as a disaster, can be straightforward due to potentially larger amounts of relevant data. However, social media data during periods without major events are likely to contain very few, if any, relevant tweets. As such, if the user \textit{only} trains the model on irrelevant data, it will poorly predict relevant data since it has only be taught what is irrelevant. Although the user can improve the model through training during a real-life disaster, they are required to know when and where the disaster occurs to begin training. This can be problematic if the user wishes to rely on relevancy predictions to detect hazardous situations. 

To accommodate time periods in which relevant data is scarce, we plan to introduce an interactive feature in which users can provide example tweets or external resources for specific relevance labels. For instance, if the user would like to train a \textit{Hurricane} classifier before a hurricane event, they could provide a relevant text such as ``I'm stranded by this hurricane. Please help!''. The model could then detect relevant tweets once the hurricane begins as opposed to requiring user training during the event. In addition, we plan to provide the user with the option to visit specific historical data to train existing classifiers, as done by Bosch et al.~\cite{BoschHarald2013SRMo}.

Our interactive learning performed well on target datasets (i.e., wildfire, bombing, and crash) as explained in Section 3.3.5. Specifically, it required the users to label approximately 200 tweets to achieve acceptable $F_1$ scores. However, tweets are short, and therefore, more research is required to investigate the suitability of our approach for ``general" classifiers, such as ones that learn to classify relevant data to ``safety". As safety can be affected by many events or situations, the model may need additional training that is typical of a targeted dataset.

Finally, although we rigorously optimize and evaluate our machine learning model,
the hyperparameter combinations were only tuned with the Figure Eight dataset~\cite{figureeight}.
Since optimal hyperparameters can depend on the dataset itself, it is possible our model may not be optimally tuned for different datasets, even though that optimization may be negligible from a user standpoint. We did use other datasets in our model evaluation to show the satisfactory resulting performance (Section 3.3.5).
Given that the Figure Eight dataset classifies generic events as relevant or irrelevant, as opposed to specific events, we expect that our model performs well on many different event types.
\section{Conclusion}
We presented a novel interactive framework in which users iteratively (re)train neural network models with streaming text data in real-time to improve the process of finding relevant information. 
We optimized and evaluated a machine learning model with various datasets related to situational awareness and adapted the model to learn at interactive rates.
According to evaluation results, our model outperforms state-of-the-art learning models used in similar classification tasks. 
Finally, we integrated our framework with the SMART application and extended it to SMART 2.0, allowing users to interactively explore, identify, and refine tweet relevancy to support real-time situational awareness.
Our discussions with multiple first responders who use SMART 2.0 indicated positive feedback and user experience.
In particular, their assessments demonstrated that our interactive framework significantly improved the time-consuming process of finding crucial information during real-time events.

\section{Acknowledgements}
This material is based upon work funded by the U.S. Department of Homeland Security (DHS) VACCINE Center under Award No. 2009-ST-061-CI0003, DHS Cooperative Agreement No. 2014-ST-061-ML0001, and DHS Science and Technology Directorate Award No. 70RSAT18CB0000004.
\bibliographystyle{abbrv-doi-narrow}
\bibliography{interactivelearning}
\end{document}